# CASA-Based Speaker Identification Using Cascaded GMM-CNN Classifier in Noisy and Emotional Talking Conditions

[1,*]Ali Bou Nassif, [2]Ismail Shahin, [3]Shibani Hamsa, [4]Nawel Nemmour, [5]Keikichi Hirose

[1]Department of Computer Eng., University of Sharjah, Sharjah, United Arab Emirates

[2,3,4]Department of Electrical Eng., University of Sharjah, Sharjah, United Arab Emirates

[5]The University of Tokyo, Tokyo, Japan

[1]anassif@sharjah.ac.ae, [2]ismail@sharjah.ac.ae, [3]shibani.h@gmail.com, [4]nemours90@hotmail.com, [5]hirose@gavo.t.u-tokyo.ac.jp

*Corresponding Author: Ali Bou Nassif

## Abstract

This work aims at intensifying text-independent speaker identification performance in real application situations such as noisy and emotional talking conditions. This is achieved by incorporating two different modules: a Computational Auditory Scene Analysis (CASA) based pre-processing module for noise reduction and "cascaded Gaussian Mixture Model – Convolutional Neural Network (GMM-CNN) classifier for speaker identification" followed by emotion recognition. This research proposes and evaluates a novel algorithm to improve the accuracy of speaker identification in emotional and highly-noise susceptible conditions. Experiments demonstrate that the proposed model yields promising results in comparison with other classifiers when "Speech Under Simulated and Actual Stress (SUSAS) database, Emirati Speech Database (ESD), the Ryerson Audio-Visual Database of Emotional Speech and Song (RAVDESS)" database and the "Fluent Speech Commands" database are used in a noisy environment.

**Keywords:** Computational Auditory Scene Analysis (CASA); Convolutional Neural Network; Gaussian Mixture Model; Speaker Identification.

## 1. Introduction and Literature Review

Research and developments in speaker recognition systems have resulted in a vast range of acceptance in several fields such as banking, forensic authorization and security applications in neutral talking condition [1],[2]. However, the system performance degrades in noisy and emotional talking conditions [3]. The challenges of designing a system which can offer a high performance in the midst of a noisy interference signal and in emotional talking conditions have been identified as the main objectives of this paper.

Speech is considered to be the primary communication system [4]. An effective communication system consists of both a linguistic and an emotional/stressful part [5]. The emotional/stressful aspect of human communication systems is inevitable in any Human-Machine interaction system. Speaker identification in emotional and noisy talking conditions can be used to offer a promising future for the banking sector, smart customer support, and forensic applications [6,7].

In natural conditions, speech undergoes various kinds of interference, such as surface reflections, reverberations, noise from other sources, and so on [8]. These interferences, which are present in the dominant speech signal, may reduce the system's performance at the application level [9]. For example, a bank security system using speaker identification may fail to work in noisy conditions. This is because the system does not have the ability to separate the dominant original signal from other noisy signals. Humans have the ability to separate the dominant signal even in the presence of noise, and this ability is referred to as Auditory Scene Analysis (ASA) [10]. ASA is accomplished by the combined efforts of the human auditory and intelligence systems. The system performance can be improved by the incorporation of computationally developed ASA for machines. Therefore, they can separate the dominant signal from other forms of interference before verification by means of Computational Auditory Scene Analysis (CASA) [11]. Our proposed model show promising results in comparison with other classifiers such as Support Vector Machine (SVM) and Multilayer Perceptron (MLP).

These days organizations, industries and several homes are equipped with security devices such as surveillance cameras. These devices can capture sounds of the surroundings. They may capture the voice of predators even in noisy and emotional conditions. The recorded voice can be an input to the proposed system, and this can be used by police in criminal investigations to identify a potential criminal for example. Four distinct speech datasets have been included in this work to assess the proposed model.

The implementation of emotion recognition, along with noise suppression is of great importance in the development of speaker identification techniques for successful implementation of an effective human-machine interaction system.

Shao and Wang [12] studied acoustic features and investigated a general solution to achieve a robust speaker identification system under noisy environments. They proposed state of the art speaker-dependent features obtained from auditory filtering and cepstral analysis. These auditory features were further enhanced, by means of binary time-frequency (T-F) masks produced by a CASA system, and their reconstruction uncertainties were estimated for better computation of speaker likelihood. Results demonstrated that their proposed Gammatone Frequency Cepstral Coefficients (GFCCs) features perform significantly better than the traditional Mel-Frequency Cepstral Coefficients (MFCC) features. Ghiurcau et al. [13] evaluated the impact of speaker emotional features upon text-independent speaker recognition systems by means of MFCCs, for feature extraction, and the SVM model, for classification. Experimental tests were performed on the Berlin emotional speech database. Results demonstrated that emotions play a vital role in minimizing the performance of speaker identification systems compared to when these same words were spoken neutrally.

Zhao et al. [14] studied speaker identification performance under noisy environments. The authors first presented a new speaker feature, called gammatone frequency cepstral coefficient (GFCC) and demonstrated that this auditory feature picks up seizures acoustic characteristics pertinent to the speaker and performs noticeably better than the conventional speaker features under noisy conditions. At a later stage, the authors applied CASA separation and then reconstructed or marginalized the deteriorated constituents, specified by the CASA mask. They found out that both reconstruction and marginalization are effective. Li et al. [15] proposed a novel architecture to enhance the robustness in emotion-dependent speaker recognition/identification systems. In fact, they proposed a new learning technology to reweight the probability of test affective utterances at the pitch envelop level. Experiments were carried out upon the Mandarin Affective Speech dataset and results yielded an enhancement of 8% of identification performance rate over the conventional speaker recognition schemes. Patnala and Prasad [16] proposed a novel scheme in order to enhance speaker identification performance under the existence of fused effects of additive noise and room reverberations, which together present a significant challenge to building robust solutions to related systems. The authors proposed a system solution with the aim of resolving the aforementioned matter using two steps. The first step was the preprocessing of the audio signal corrupted by noise and room reverberations using binary time-frequency (T-F) masking algorithm, using a CASA approach, via a deep neural network classifier. Mansour et al. [17] employed the i-vector approach along with the Support Vector Machine (SVM) classifier as an attempt to boost and enhance the deteriorated performance of speaker recognition under emotional auditory environments. Results showed that the i-vector algorithm resolves the problem of training algorithm complexity that the SVM model suffers from and shows promising results in increasing speaker recognition performance in an emotional context. Islam et al. [18] proposed a state of the art neural-response-based method for a speaker identification system under noisy acoustic conditions using 2-D neurograms coefficients, which are coefficients built upon reactions of a physiologically-based computational model of the auditory periphery. The classification accuracies of the proposed model were compared to the performances of the traditional speaker identification methodologies using features such as "MFCCs, GFCC) and Frequency domain linear prediction (FDLP)". The identification results attained by the proposed method were comparable to the performance of those conventional approaches in quiet settings, but the new feature has demonstrated lower classification error rates under noisy environments.

Faragallah [19] advocated a speaker identification system, that is resistant to noise, named MKMFCC–SVM. This system is based on the "Multiple Kernel Weighted Mel Frequency Cepstral Coefficient (MKMFCC) and support vector machine (SVM)". A comparison was made between the performance of the proposed "MKMFCC–SVM and the MFCC–SVM" speaker identification systems. Results revealed that the proposed MKMFCC–SVM system produces better identification rates in the presence of noise. Korba et al. [20] stated that MFCC features are deemed very sensitive in the presence of background conditions, which has a considerable negative impact on the performance of speaker identification systems. The authors combined the features they obtained with MFCC features. Their speaker identification system was implemented on the GMM using TIMIT speech corpus. The results of their method of implementation and testing were increased up to 28% accuracy at signal to noise ratio (SNR) 5 dB. Ayhan and Kwan [21] developed a vigorous speaker identification scheme under noisy conditions which implicates "mask estimation, gammatone features with bounded marginalization to deal with unreliable features, and Gaussian mixture model (GMM) for speaker identification". Evaluation and assessments were performed to determine the speaker identification performance of the proposed algorithm, and results showed that it substantially outperforms the conventional method MFCC with Cepstral Mean Normalization (MFCC-CMN) at low signal-to-noise conditions. Nasr et al. [22] proposed a new framework to enhance speaker identification accuracy based on "Cepstral features and the Normalized Pitch Frequency (NPF)". The novel approach used a neural classifier with a single hidden layer node as well as a pre-processing noise reduction step prior to the feature extraction procedure in order to enlarge and enhance the identification performance.

There are several limitations in the related work. Much of the literature on this subject attempted to propose groundbreaking approaches and pioneering methodologies with the aim of enhancing speaker identification accuracy under noisy as well as emotional environments. Some authors used the conventional MFCC features [13],[23]; while some others introduced novel acoustic features such as GFCCs features [12] and 2-D neurograms coefficients [18]. Some scholars favored examining the use of CASA modules in noisy speech, in conjunction with one of the above-mentioned acoustic features, and results showed substantial improvement in identification performance in some cases. Moreover, many studies used the conventional classifiers, such as SVMs [13], GMMs [18],[21] and HMMs [23–25], while many recent work explored the DNN-based classifiers [16],[26].

The aim of this study is to introduce a novel algorithm for speaker identification in real-world applications. Speech processing modules are susceptible to noise and interference in natural environments. This reduces the system's performance in real world applications. In contrast, the proposed algorithm can identify the speaker in noisy and emotional talking conditions. The proposed algorithm incorporates a CASA pre-processing module for noise suppression and cascaded GMM-CNN classifier for emotion recognition and speaker identification.

To the best of our knowledge, none of the former studies has considered the usage of CASA pre-processing module and MFCC-based feature extraction in combination with hybrid cascaded DNN-based classifier, such as GMM-CNN classifier, in order to boost text-independent speaker identification systems under noisy and emotional talking conditions.

Our contributions in this work are:

- To the best of our knowledge, this is the first work that proposes CASA-GMM-CNN model.
- Implementation of emotion recognition, by means of the GMM model; along with the CNN, for final identification decisions, which results in a hybrid GMM-CNN classification model.
- Implementation of the CASA pre-processing method and the MFCC based feature extraction together with the hybrid cascaded classifier, GMM-CNN.
- The proposed framework is capable of separating the original speech signal from other noise and interference.
- The proposed system is able to recognize the unknown speaker even in emotional/stressful talking conditions.

The remainder of the paper is structured as follows: Sections 2 presents the materials and methods used in this research. Section 3 depicts the results and provides a discussion about the results. Finally, Section 4 concludes our work.

## 2. Materials and Methods

### a. Speech Databases

In this work, four distinct datasets have been utilized to evaluate the proposed model. The datasets are Speech Under Simulated and Actual Stress (SUSAS) dataset [27], Arabic private Emirati Speech Database and the Ryerson Audio-Visual Database of Emotional Speech and Song (RAVDESS) [28]. The four databases are listed as follows:

**Speech Under Simulated and Actual Stress (SUSAS) dataset**

SUSAS is an English public dataset that consists of five domains which have an array of stress and emotion features [27]. The database has two domains; one involves simulated speech under stress and is termed Simulated Domain. The second involves actual speech under stress and is termed Actual Domain. A group of thirty-two speakers including 19 males and 13 females, in the age group 22 to 76 years, were asked to pronounce more than 16,000 words. The speech tokens were sampled at a frequency of 8 kHz using 16 bits A/D converter. The signal samples were pre-emphasized and then segmented into frames of 20 ms each with 31.25% overlap between consecutive frames. The emphasized speech signals were implemented every 5 ms to a 30 ms Hamming Window. The observation vectors in each of CASA-based GMM-CNN were found using a 32-dimension feature analysis of MFCCs (sixteen delta MFCCs and sixteen static MFCCs). In this work, twenty different words, uttered twice by twenty speakers (two repetitions per word), uttered in seven stressful talking conditions were used. Out of the twenty words, ten words were used for training and twenty for testing. During the evaluation phase, ten different words were uttered by ten speakers twenty-five times under seven stressful talking conditions, which are neutral, angry, slow, loud, soft, Lombard and fast. These were mixed with the other speech signals in the same database in a ratio of 2:1 and 3:1 and were then used. Ten different words uttered by same ten speakers two times under six stressful talking conditions were mixed with various noise signals in the ratio 2:1 and 3:1.

**Arabic Emirati Speech Database (ESD)**

ESD is a private dataset made up of Twenty-five female and twenty-five male Emirati speakers with age range spanning from fourteen to fifty-five years old articulated the "Arabic Emirati-emphasized speech database". Eight common Emirati utterances, frequently used in the United Arab Emirates society, were uttered by every speaker. Every speaker expressed the eight sentences in each of neutral, happy, sad, disgusted, angry, and fearful emotions, nine times with a span of 2 – 5 seconds. The captured dataset was recorded in the "College of Communication, University of Sharjah, United Arab

Emirates". During the training stage, the first four sentences were used, while in the testing phase, the remaining four utterances were utilized. The database was collected by a speech acquisition board using a 16-bit linear coding A/D converter and sampled at a sampling rate of 44.6 kHz. The signals were then down sampled to 12 kHz. The samples of signals were pre-emphasized and then segmented into slices (frames) of 20 ms each with 31.25% intersection between successive slices". The emphasized speech signals were applied every 5 ms to a 30 ms Hamming Window.

**The Ryerson Audio-Visual Database of Emotional Speech and Song (RAVDESS)**

RAVDESS is a public English dataset that has been used to assess the proposed model [28]. The RAVDESS consists of 24 professional speakers (12 males and 12 females), expressing two lexically matched speeches in a neutral North American accent. RAVDESS has two spoken statements: "Kids are talking by the door" and "Dogs are sitting by the door". Speech emotions contain neutral, angry, happy, sad, fear, and disgust emotions. RAVDESS contains 7356 files (Audio and Audio -visual). In this work we have used a total of 2452 files. Two lexically matched statements were spoken by every speaker in 60 trials constituting 1440 speech files (60 attempts per speaker × 24 speakers) and 44 trials of twenty-three speakers contributes 1012 song files are used in this research.

**Fluent Speech Commands**

The Fluent Speech Commands dataset [29] contains 30,043 utterances from 97 speakers. It is recorded as 16 kHz single-channel .wav files each containing a single utterance used for controlling smart-home appliances or virtual assistant. The dataset has a total of 248 phrasing mapping to 31 unique intents. The utterances are randomly divided into train, valid, and test splits in such a way that no speaker appears in more than one split. Each split contains all possible wordings for each intent.

## b. CASA Pre-Processing for Noise Reduction

The proposed system incorporates a CASA-based preprocessing module for co-channel noise reduction. Figure 1 shows the CASA based speech separation block diagram. This figure consists of modulation frequency analysis, smoothing, onset-offset detection, segmentation and grouping [30].

**T-F Decomposition**

The speech signal that needs to be identified, is broken up into small time frame signals for segmental feature extraction and processing [31]. Time-frequency (T-F) analysis of each time frame is computed by taking its short-time Fourier Transform (STFT) and is recorded as a matrix which can track the magnitude and phase in time–frequency domain [32].

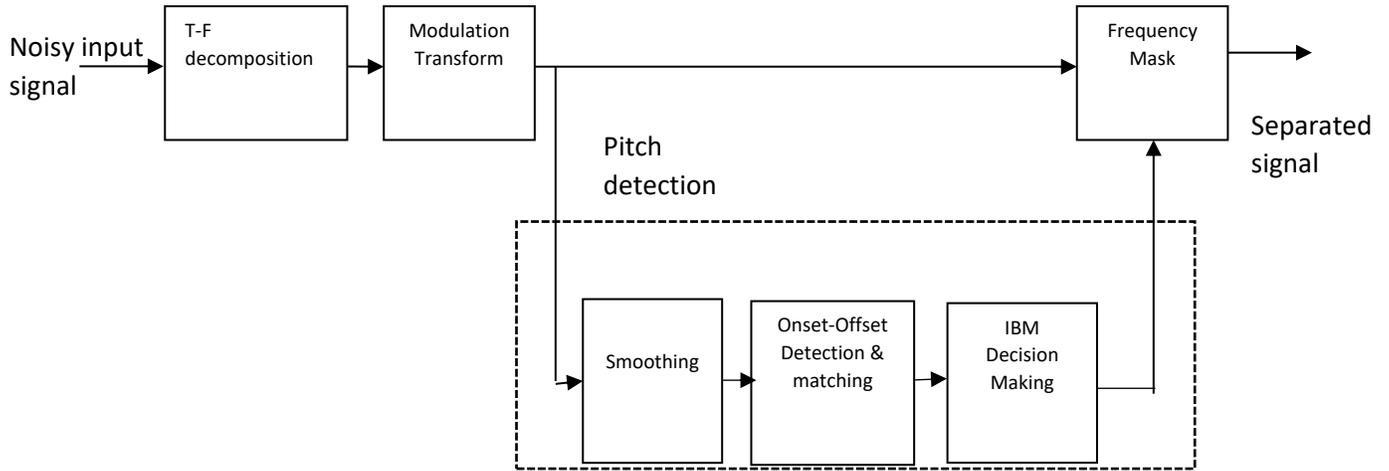

Figure 1 Original speech separation from noisy input signal

**Modulation Transform**

The signal $X(m, k)$ extracted by the T-F decomposition process consists of the Modulator Signal $M(m, k)$ and the Carrier Signal $C(m, k)$ [32]. This can be defined as,

$$X(m, k) = M(m, k)\, C(m, k) \tag{1}$$

The modulator of the signal $X(m, k)$, $M(m, k)$, can be obtained through applying an envelope detector. It can be represented as,

$$M(m, k) \cong ev\{X(m, k)\} \tag{2}$$

where "ev" denotes the envelope detection. The envelope detector used is an incoherent detector which is based on "Hilbert" envelope [33] as it is able to create a modulation spectrum with large area covered in the modulation frequency domain. In addition, it acts as a magnitude operator for complex-valued sub-bands, as given by the following equation,

$$M(m, k) \cong |X(m, k)| \tag{3}$$

Then, the Discrete Short-time Modulation Transform of the signal $x(n)$ can be expressed as,

$$X(k, i) = DFT\{D\{ev\{STFT\{x(n)\}\}\}\}$$
$$= \sum_{m=0}^{I-1} M(m, k)\, e^{-j2\pi mi/I} \quad i = 0, \ldots, I\text{-}1 \tag{4}$$

where I is the DFT length and i represents the modulation frequency index. $ev\{STFT\{x(n)\}\}$, is the modulating signal part and it is denoted as $M(m, k)$.

**Onset-Offset Position Analysis**

Many of the speech separation or noise reduction techniques using the CASA algorithm are performing some kind of speech enhancement or noise reduction. Using a low pass filter, the modulation transformed signal is smoothed. The partial derivative of this signal with respect to its modulating frequency will aid in the identification of the peaks and valleys of the signal which can be termed onset position and offset position, respectively [32].

**Binary Mask Segmentation**

The onset-offset positions extracted from the likely originated sources are to be grouped to form segments. This can be accomplished by means of an "Ideal Binary Mask" (IBM), which can be expressed as [34],[35],

$$\text{IBM }(t,f) = 1; \text{ if } f_d = f_{on} + \frac{\rho f_s}{N} \tag{5}$$

where $f_d$ is the dominant pitch frequency of the input signal computed by autocorrelation based pitch estimation [36], fs is the sampling frequency and $\rho$ varies from -10 to 10.

Then, the masked signal can be denoted as [32],

$$S_{IBM}(t,f) = \begin{Bmatrix} s(t,f) & , \text{if } f = f_{on} \\ 0 & , \text{else} \end{Bmatrix} \tag{6}$$

The spectral energy of the dominant signal can be extracted from $S_{IBM}$ (t,f) and the range of the pitch of the interference is calculated from the remaining part of the mixture. Spectral energy from the dominant and interference signals in the entire pitch range can be used to design a frequency mask for separating the desired speech signal from the noise signals [37].

**Segregation Mask**

The speech signal can be segregated by means of a frequency based separation mask. The input signal x(n) sampled at a rate of $f_s$ consists of both speech signal $x_n$ (n) and interference signal $x_t$ (n) as,

$$x(n) = x_t(n) + x_n(n) \tag{7}$$

The mean of the signal spectral energy of the speech and noise signals are estimated for designing a suitable frequency mask for noise suppression. $X_T$ (k) is the mean modulation spectral energy over the pitch frequency of the target signal and $X_I$ (k) is the mean modulation spectral energy over the pitch frequency of the interference signal [38],

$$X_T(k) = \frac{\sum_J |S(m,k)|^2}{\text{Pitch frequency range of dominent speech signal}} \tag{8}$$

$$X_I(k) = \frac{\sum_J |S(m,k)|^2}{\text{pitch frequency range of inteference signal}} \tag{9}$$

Frequency mask can be designed as,

$$F(k,i) = \frac{X_T(k)}{[X_T(k) + X_I(k)]} \tag{10}$$

### c. Features Extraction

The short term power spectrum of the sound can be effectively represented by Mel Frequency Cepstral Coefficients (MFCCs) [39]. In Mel Frequency Cepstrum (MFC), filter coefficients are equally spaced in mel scale rather than linearly spaced filter coefficients in the normal scale. Hence, MFC can efficiently represent the human sound signals accurately [35],[40],[41].

The periodogram-based power spectral estimate of the target speech frame $s_t(m, k)$ for the $m^{th}$ frame at the $k^{th}$ frequency bin index can be expressed as follows [42],[43],

$$P_i(k) = \frac{1}{N} |s_t(m, k)|^2 = \frac{1}{N} \left| \sum_{n=0}^{N-1} x(n) e^{-j2\pi fn} \right|^2 \tag{11}$$

where k and N represents the index of the frequency bin, k = 0, .., K-1 and the frame length, respectively.

In order to compute the mel-spaced filter bank, the sum of the periodogram power spectral estimate of 26 triangular filters are calculated. Log value of each of the 26 energies will give log filter bank energies. Discrete Cosine Transform (DCT) of the log filter bank energies are computed to get MFCC [42],[43].

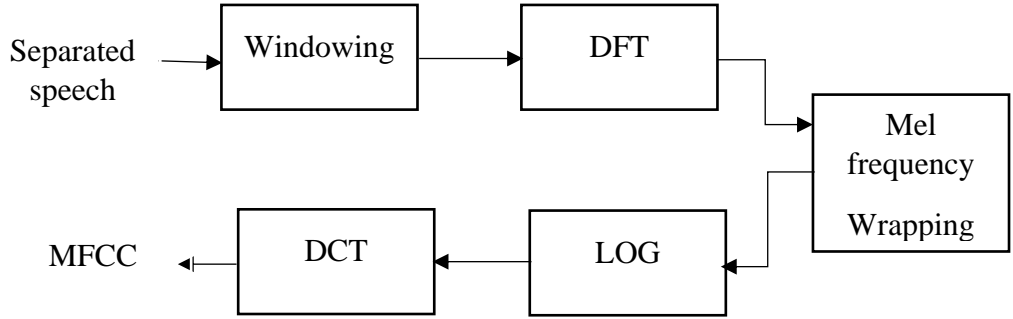

Figure 2 MFCC feature extraction [39]

Figure 2 shows the basic flow diagram of MFCC extraction. In this figure, windowing involves the slicing of the audio waveform into sliding frames using Hamming window. The chopped frame with Hamming window maintains the original frequency information better with less noise compared to a rectangle window. Then, DFT is applied to extract information in the frequency domain. In feature extraction, triangular band-pass filters are used to convert the frequency information to mimic what a human perceived. The next step is to "calculate the power spectrum of each frame". This is motivated by the human cochlea, which vibrates at different spots depending on the frequency of the incoming sounds. The periodogram spectral estimate still contains a lot of information not required for speaker identification. For this reason, we take clumps of periodogram bins and sum them up to get an idea of how much energy exists in various frequency regions. This is performed by means of Mel filterbank. The Mel scale estimates exactly how to space our filterbanks and how wide to make them. Then, the obtained filterbank energies are converted to log scale for channel normalization. Finally, Discrete

Cosine Transform (DCT) of the 26 log filterbank energies are computed to give 26 cepstral coefficients."

### d. Proposed Design

Speaker recognition and emotion recognition use a variety of classifiers such as SVM, K-NN, GMM, HMM and Multilayer Perceptron (MLP). Among all these classifiers, many studies in speech processing used the GMM classifier since it can represent the complex distribution attributes as multiple modes [43]. Hence, the GMM classifier is considered as a suitable selection for noise suppression, speaker identification, and emotion recognition applications.

**GMM Tag Generation**

The Gaussian mixture density model is a weighted sum of *M* component densities and it can be defined as [43],[44],

$$P(\bar{x}|\lambda) = \sum_{i=1}^{M} P_i b_i(\bar{x}) \tag{12}$$

where $\bar{x}$ and $b_i(\bar{x})$ represent the D-dimensional random vector and the component densities for i = 1, . M, respectively. The $P_i$, for i = 1,..,M, are the mixture weights. The $b_i(\bar{x})$ is given by the equation,

$$b_i(\bar{x}) = \frac{1}{(2\pi)^{D/2}|\Sigma_i|^{1/2}} \exp\left\{\frac{-1}{2}(\bar{x} - \bar{\mu}_i)' \Sigma_i^{-1}(\bar{x} - \bar{\mu}_i)\right\} \tag{13}$$

where $\bar{\mu}_i$ and $\Sigma$ i are the mean vector and the covariance matrix, respectively.

The GMM tag λ is the collective representation of the following GMM parameters: mean $\bar{\mu}_i$, covariance $\Sigma_i$, and weights $P_i$. It is expressed by the following notation,

$$\lambda = \{P_i, \bar{\mu}_i, \Sigma_i\}, \text{ for } i = 1,.., M \tag{14}$$

**GMM Evaluation**

The "speaker identification" algorithm which is based on the GMM classifier uses the features extracted from the test signal. After that, the complex feature distribution is converted into multiple modes of length *T*. The algorithm uses a convergence method as explained below [44]:
1. The training of the GMM model is initialized with λ
2. The new model $\bar{\lambda}$, is computed. Thereby, $p(X|\bar{\lambda}) \geq p(X|\lambda)$.
3. The process is repeated until the convergence is achieved,

$$p(i|\vec{x}_t, \lambda) = \frac{p_i b_i(\vec{x}_t)}{\sum_{k=1}^{M} p_k b_k(\vec{x}_t)} \tag{15}$$

Mixture weights are termed as,

$$\bar{p}_i = \frac{1}{T} \sum_{t=1}^{T} p(i|\vec{x}_t, \lambda) \tag{16}$$

Means are given by,

$$\vec{\bar{\mu}}_i = \frac{\sum_{t=1}^{T} p(i|\vec{x}_t, \lambda) \vec{x}_t}{\sum_{t=1}^{T} p(i|\vec{x}_t, \lambda)} \quad (17)$$

Variance is defined as,

$$\bar{\sigma}_i^2 = \frac{\sum_{t=1}^{T} p(i|\vec{x}_t, \lambda) x_t^2}{\sum_{t=1}^{T} p(i|\vec{x}_t, \lambda)} - \bar{\mu}_i^2 \quad (18)$$

where $\sigma_i^2$, $x_t$ and $\mu_i$ are arbitrary elements of the vectors $\bar{\sigma}_1^2, \bar{x}_t$ and $\bar{\mu}_1$, respectively.

Speaker set S = {1, 2, 3, …, s} is denoted by GMM's $\lambda_1, \lambda_2, \ldots, \lambda_s$. The speaker model is defined as,

$$\hat{S} = \arg\max_{1 \leq k \leq S} \sum_{t=1}^{T} \log P(\bar{x}_t | \lambda_k) \quad (19)$$

in which $P(\bar{x}_t | \lambda_k)$ is given in (14).

**CNN Classifier**

Convolutional Neural Networks (CNN) classification is one of the cutting-edge classification techniques in machine learning [45],[46],[47],[48]. In deep learning, CNN models are part of deep neural networks (DNN). CNN classifiers are applicable in acoustic signal processing, as well as other applications. The CNN classifier is used for the precise identification of the target speaker, followed by the GMM classification. A 50-layer Convolutional Neural Network (CNN) is employed for classification. Each convolutional layer is followed by a maxpooling layer. The fully connected layers use GMM tags to tune the final result from the CNN classifier. For every input at the fully connected layers, the system evaluates the GMM tag value in order to filter the results at the output stage. The decision will be a binary 0 or 1 based on the GMM tag.

This paper uses a CNN with fifty hidden layers in addition to input and output layers. The activation function used in the hidden neurons is the "Rectified Linear Unit ReLU activation [49]. After training, the "CNN model" produces probability distribution *P* over all emotions. After that, the decision block selects the particular model having the highest probability value." The speech signal consists of linguistic part, emotional/stressful part, noise and distortions. Hence, the simple speaker identification system with feature extraction followed by classification is not sufficient to support human-machine interaction systems. This work proposes an efficient speaker identification algorithm that can identify and recognize the speaker in both emotional and noisy talking conditions. This is achieved by incorporating CASA-based pre-processing module, MFCC based feature extraction and cascaded GMM-CNN classifier.

Figure 3 demonstrates the basic schematic blocks of the proposed speaker identification system. It consists of CASA pre-processing module, feature extraction using MFCC and classification based on cascaded GMM-CNN classifier. The CASA system receives the input signal s(n) from noisy emotional talking conditions. S(m,k) is the T-F transformed narrow band signal. Envelop detection over S(m,k) gives the modulating signal M(m,k). The next step is to form the segments by matching individual onset and offset values by using Ideal Binary Mask. Two almost disjoint segments with the most modulation spectrogram energies are used for the generation of frequency mask. The target dominant signal can be obtained by taking the convolution of the modulating signal and frequency filter impulse response.

The noise of the target signal features is extracted using MFCC and are fed to the cascaded "GMM-CNN classifier. During the evaluation phase, the log likelihood distance between the voice query and each of the GMM tags is compared for each of the emotional/stressful state and, thus, produces a recent vector of features, which is considered as the input of the CNN classifier." The CNN classifier provides the final decision.

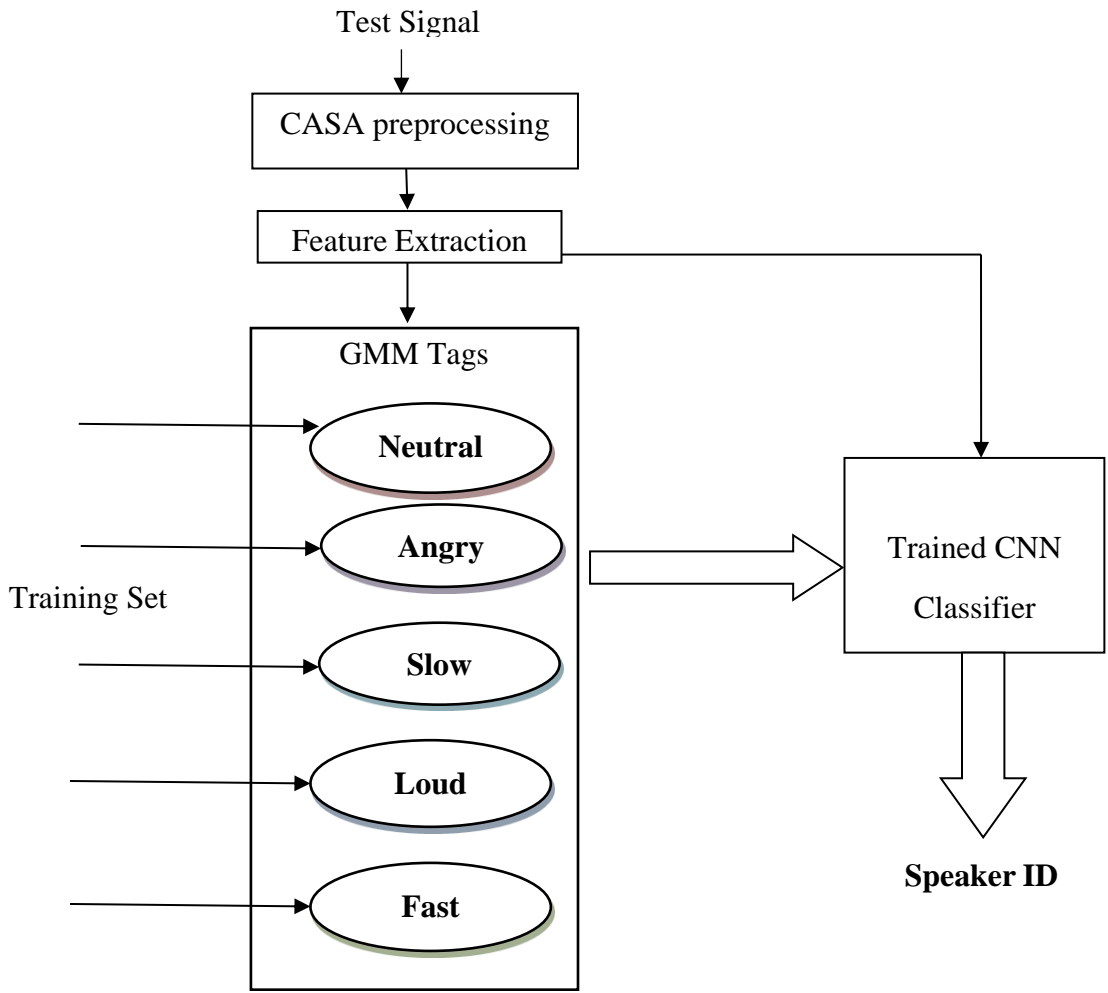

Figure 3 Block schematic of the proposed system

The algorithm of the whole system is shown below:

1. The training of the GMM model is initialized with λ
2. The new model $\bar{\lambda}$, is computed. Thereby, $p(X|\bar{\lambda}) \geq p(X|\lambda)$.
3. The process is repeated until the convergence is achieved,
4. Select the GMM tags with maximum probability of convergence
5. Selected GMM tags are used as the input of CNN classifier
6. Fine tuning using CNN decides the final decision.

## 3. Results and Discussion

This work implements the proposed GMM-CNN model for effective speaker identification in emotional and noisy talking conditions. To evaluate the proposed algorithm, evaluation metrics such as Speaker Identification Performance (SID), Precision, Recall, F1 score and Area Under the Curve (AUC) metrics have been used.

The results show that every model functions almost ideally in neutral talking conditions. The proposed GMM-CNN model outperforms all other models using the SUSAS dataset and based on the performance evaluation metrics reported in Equations 20 to 23 [50]:

$$\text{SID Performance} = \frac{\text{"Total number of times the unknown speaker has been identified correctly"}}{\text{"Total number of trials"}} \times 100\% \quad (20)$$

$$Precision = \frac{TP}{TP+FP} \quad (21)$$

$$Recall = \frac{TP}{TP+FN} \quad (22)$$

$$F1\ Score = \frac{Recall*Precision}{Recall+Precision} \quad (23)$$

Where "TP, TN, FP and FN are the True Positive, True Negative, False Positive and False Negative values, respectively are obtained from the confusion matrix."

The average text-independent speaker identification in each of neutral and emotional/stressful environments using the SUSAS dataset in view of each of CASA-based GMM-CNN, GMM-CNN, SVM and MLP is 84.49%, 80.45%, 76.77% and 77.24%, respectively as illustrated in Figure 4. This shows that the CASA-based GMM-CNN model outperforms other models using SUSAS database. Moreover, the highest and lowest SIDS are reported for Neutral and Angry, respectively and this is consistent with prior work. In order to confirm our results, statistical tests should be used to check if the CASA-based GMM-CNN is statistically different from other models. Before we use a proper statistical test, we have to check the distribution of the SID Performance. Using the Kolmogorov-Smirnov normality test, we found that SID Performance is not normally distributed, so we have to use non-parametric tests [51]. The Wilcoxon test [51], which is a non-parametric test was used to compare two models. Based on the results, we notice that the proposed CASA-based GMM-CNN is statistically different from other models based on 90% confidence level. Hence, we can confirm that the CASA-based GMM-CNN model surpasses other models and it is also statistically different from other models.

In order to generalize the validity of the proposed model, we conducted six additional experiments to assess the speaker identification performance achieved in neutral and emotional/stressful conditions using the CASA-based GMM-CNN classification algorithm. These experiments are:

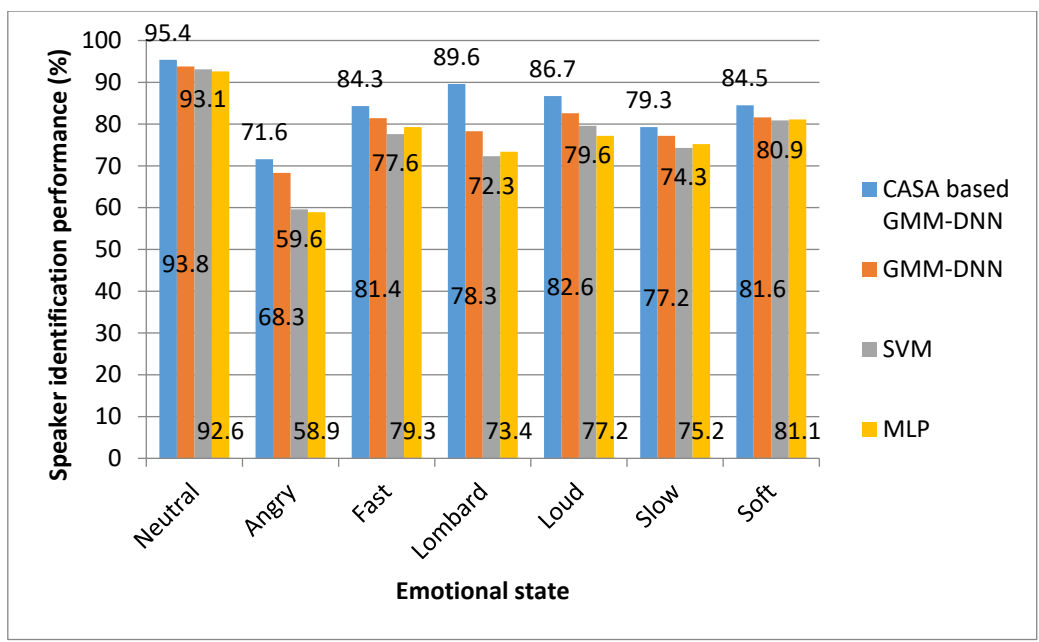

Figure 4 "Average speaker identification performance evaluation" based on CASA-based GMM-DNN, GMM-DNN, SVM and MLP using SUSAS database

**Experiment 1:** The proposed CASA-based cascaded GMM-CNN classification algorithm for speaker identification in noisy and emotional talking conditions is evaluated by comparing it with other classifiers. This can be achieved by adding CASA preprocessing module to the SVM and MLP classifiers. Based on Figure 5 and Table 1, the proposed CASA-based cascaded GMM-CNN classifier shows significant improvement over the CASA-based SVM and CASA-based MLP.

**Table 1**

Evaluation of GMM-CNN, SVM and MLP utilizing SUSAS database

|  | GMM-CNN | SVM | MLP |
|---|---|---|---|
| **SID** | 0.84 | 0.78 | 0.78 |
| **Precision** | 0.80 | 0.70 | 0.70 |
| **Recall** | 0.82 | 0.72 | 0.69 |
| **F1 score** | 0.81 | 0.71 | 0.69 |
| **ROC** | 0.80 | 0.70 | 0.60 |

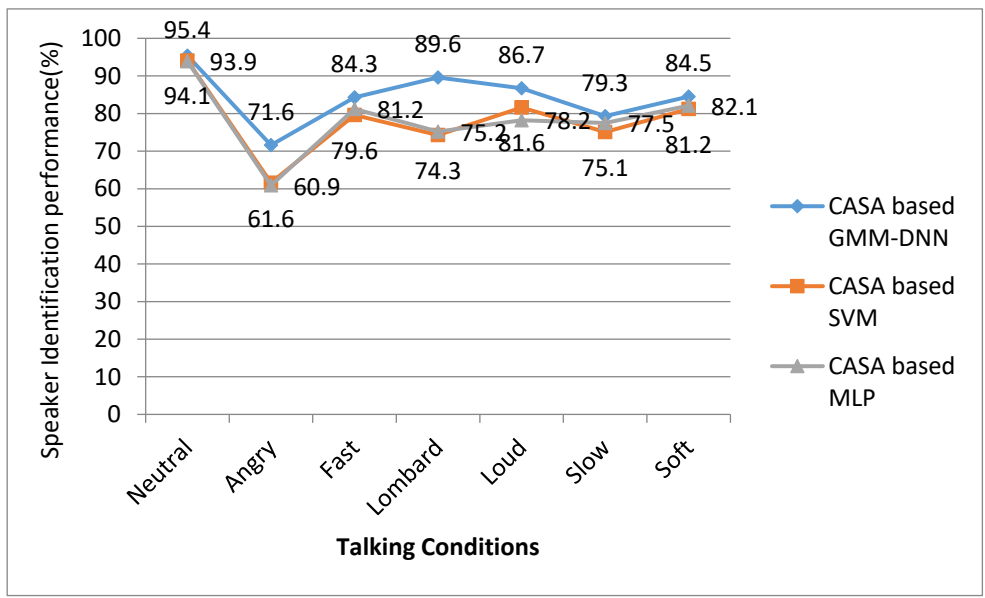

Figure 5 Average speaker identification performance evaluation using SUSAS based on each of CASA-based GMM-CNN, CASA-based SVM and CASA-based MLP

**Experiment 2**: Another assessment of the proposed CASA-based GMM-CNN using the SUSAS database has been conducted using ten nonprofessional audience members (human judges). Overall, thirty speech samples are used in this experiment. During the testing stage, the speech samples were mixed with noise signals in a ratio 2:1. Figure 6 illustrates that the "human listener performance" is close to the proposed CASA-based GMM-CNN system.

**Experiment 3:** The proposed system was also evaluated by using a private Arabic Emirati-accented dataset. In this experiment, a "32-dimension feature analysis of MFCCs (16 static MFCCs and 16 delta MFCCs) was utilized to find the observation vectors in CASA-based GMM-CNN".

Table 2 reports the "average speaker identification performance" based on GMM-CNN, SVM and MLP classifiers with and without a CASA module for noise suppression using the ESD. The speaker identification algorithm with noise suppression module gives 83.68%, 75.8% and 76.50%, based on GMM-CNN, SVM, and MLP classifiers, respectively. The results indicate that the proposed CASA-based GMM-CNN classifier surpasses other classifiers in noisy environments using the ESD.

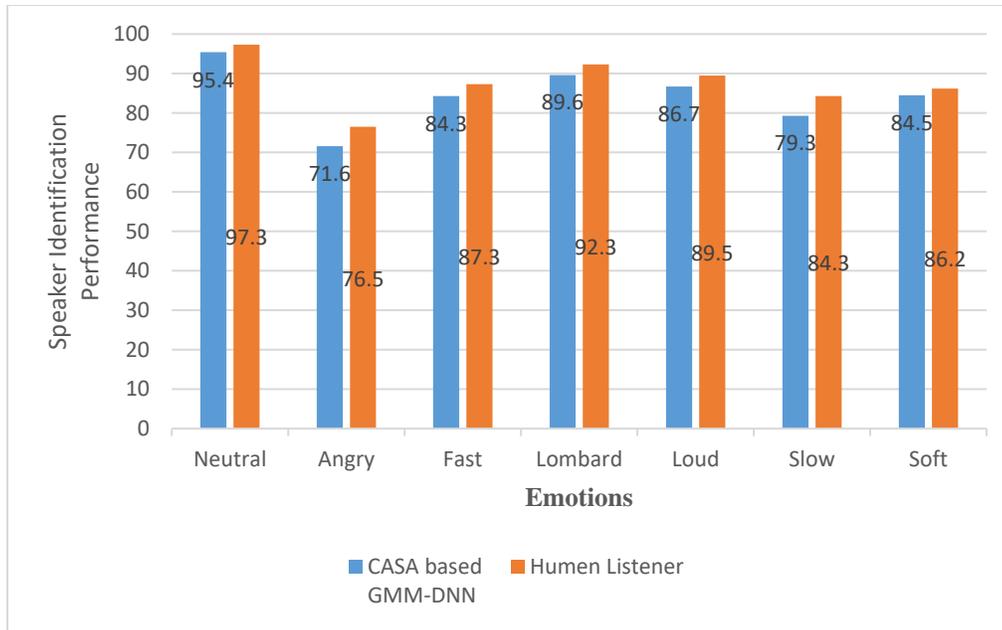

Figure 6 "Speaker identification performance analysis based on the proposed-CASA based GMM-CNN and human listeners"

Table 3 shows the average values of Precision, Recall, F1 Score and ROC (AUC) of all talking conditions based on GMM-CNN, SVM and MLP in neutral and emotional/stressful talking environments utilizing the ESD database. Based on this table, the proposed model outperforms the other models using SID, Precision, Recall, F1 Score and ROC.

**Table 2**

Speaker identification performance based on each of GMM-CNN, SVM and MLP in neutral and emotional/stressful talking environments utilizing the ESD with and without CASA

|  | "Average Speaker Identification Performance" (%) | | | | | |
|---|---|---|---|---|---|---|
|  | "GMM-CNN" | | "SVM" | | "MLP" | |
|  | without CASA | with CASA | without CASA | with CASA | without CASA | with CASA |
| **Emotion Average** | **76.8** | **83.7** | **63.0** | **75.8** | **62.7** | **76.5** |

**Table 3**

Evaluation of GMM-CNN, SVM and MLP in neutral and emotional/stressful talking environments utilizing the ESD

| Metric | GMM-CNN | SVM | MLP |
|---|---|---|---|
| **Precision** | 0.81 | 0.74 | 0.70 |
| **Recall** | 0.80 | 0.72 | 0.69 |
| **F1 score** | 0.80 | 0.73 | 0.69 |
| **ROC** | 0.80 | 0.70 | 0.60 |

**Experiment 4:** Another assessment of the proposed CASA-based GMM-CNN has been conducted using the Ryerson Audio-Visual Database of Emotional Speech and Song (RAVDESS) dataset. This experiment evaluates the proposed system performance using angry, happy, neutral, sad, fearful, disgust, calm and surprise talking conditions using the RAVDESS dataset. Table 4 shows the average speaker identification performance based on each of GMM-CNN, SVM and MLP in neutral and emotional/stressful talking environments utilizing the RAVDESS database with and without CASA. The proposed GMM-CNN model outperforms other models with and without CASA. We also notice that the performance of the models improves when CASA is used. Table 5 shows the average values of Precision, Recall, F1 Score and ROC of all emotions of GMM-CNN, SVM and MLP in neutral and emotional/stressful talking environments utilizing the RAVDESS database.

**Table 4**

Speaker identification performance based on each of GMM-CNN, SVM and MLP in neutral and emotional/stressful talking environments utilizing the RAVDESS database with and without CASA

| | "Speaker Identification Performance" (%) | | | | | |
|---|---|---|---|---|---|---|
| | "GMM-CNN" | | "SVM" | | "MLP" | |
| | without CASA | with CASA | without CASA | with CASA | without CASA | with CASA |
| **Emotion Average** | 78.67 | 84.68 | 65.6 | 75.78 | 62.89 | 74.4 |

**Table 5**

Precision, Recall, F1 Score and ROC metrics based on each of GMM-CNN, SVM and MLP in neutral and emotional/stressful talking environments utilizing the RAVDESS database.

| Metric | GMM-CNN | SVM | MLP |
|---|---|---|---|
| **Precision** | 0.82 | 0.64 | 0.67 |
| **Recall** | 0.81 | 0.62 | 0.65 |
| **F1 score** | 0.81 | 0.62 | 0.65 |
| **ROC** | 0.80 | 0.61 | 0.61 |

**Experiment 5:** Proposed CASA-based GMM-CNN performance has been evaluated using a non-emotional speech corpus called Fluent Speech Command Dataset [29]. Table 6 shows the evaluation metrics of the proposed model, as well as other models. This also confirms that the proposed model surpasses other models using this dataset.

**Table 6**

Evaluation based on each of GMM-CNN, SVM and MLP utilizing the Fluent Speech Command database.

| Metric | GMM-CNN | SVM | MLP |
|---|---|---|---|
| **SID** | 0.89 | 0.83 | 0.85 |
| **Precision** | 0.82 | 0.64 | 0.67 |
| **Recall** | 0.81 | 0.62 | 0.65 |
| **F1 score** | 0.81 | 0.62 | 0.65 |
| **ROC** | 0.80 | 0.61 | 0.61 |

**Experiment 6:** This experiment evaluates the classifiers GMM, CNN, GMM-CNN and CNN-GMM using ESD. Table 7 shows the average emotion recognition rate obtained using GMM alone, CNN alone, GMM-CNN and CNN-GMM. It is clear that the proposed GMM-CNN outperforms other classifiers. The ratio of the computational complexity with reference to GMM alone is 2, 6 and 7 respectively for CNN alone, GMM-CNN and CNN-GMM classifiers. It is evident from this experiment that hybrid classifier of GMM-CNN performs well in terms of performance with reduced computational complexity.

**Table 7**

Evaluation of GMM, CNN, GMM-CNN and CNN-GMM using ESD

|  | GMM | CNN | GMM-CNN | CNN-GMM |
|---|---|---|---|---|
| SIS | 0.71 | 0.82 | 0.87 | 0.83 |
| Complexity | 1 | 2 | 6 | 7 |

Table 8 illustrates the rate of improvement of our proposed hybrid GMM-CNN classifier over the reviewed literature. The comparison is accomplished between the highest performance achieved by GMM-CNN, among the four datasets in the noisy environment, and the average performance attained in prior work. The GMM-CNN recorded an identification performance equivalent to 84.68% for the RAVDESS database. Based on the results in Table 6, the proposed classification method demonstrates a positive improvement rate over the literature.

## 4. Conclusions

Novel CASA-based GMM-CNN classifier has been introduced and evaluated to improve the performance of text-independent speaker identification in noisy emotional talking environments using four diverse corpora.

In this work, we show that the proposed CASA GMM-CNN model has higher SID, Precision, Recall, F1 Score and ROC than that of other classifiers such as SVM and MLP. All models are evaluated using four distinct datasets including SUSAS public English dataset, ESD private Arabic dataset, RAVDESS public English dataset and the Fluent Speech Command public English dataset.

The proposed system also yields higher performance in noisy speech signals. The algorithm based on "GMM tag based-feature vector reduction" helps to minimize the complications of the CNN classifier, thus, improving system performance with reduced computational complexity. The proposed classifier outperforms other classifiers even in the presence of interference. The performance of all models has been improved when CASA system is being used.

CASA based pre-processing module makes the system more efficient in noisy talking conditions. The CASA preprocessing module segregates the dominant signal from other interference signals before performing the speaker recognition task. This leads the system to perform more efficiently even in noise susceptible real applications.

The proposed system demonstrates improvement in angry talking condition. This is achieved by the combined effects of CASA and GMM-CNN classifier systems. CASA separates the dominant signal features from the distorted input signal, which enables the classifier to perform more efficiently in such a talking condition.

The CASA based pre-processing module plays an important role in system performance. The proposed algorithm uses a STFT-based frequency mask for speech separation from the noise signal. However, there is a dilemma in Time and Frequency analysis. Larger window size offers higher accuracy in the frequency domain. Smaller window size offers better accuracy in the time domain. Accuracy in both time and frequency domains is necessary to achieve better system performance.

Further study is necessary to improve system performance. The pitch estimation method needs to be enhanced since pitch is the main cue for speech segregation and can incorporate additional preprocessing speech de-reverberation techniques to enhance the scalability in reverberant conditions.

**Table 8**

Comparison between the proposed CASA GMM-CNN classifier and related work in noisy environments (SID of CASA GMM-CNN = 84.68% for the RAVDESS database)

| Reference | Talking Environment | Classifier | Features | Dataset | Improvement Rate of the proposed model over prior work (%) |
|---|---|---|---|---|---|
| Zhao et al. [14] | noisy | GMM-UBM | GFCC coefficients | 2002 NIST Speaker Recognition Evaluation corpus | 4.33 |
| Patnala and Prasad [16] | noisy | GMM-UBM | GFCC | Private dataset | 39.08 |
| Islam et al. [18] | noisy | GMM-UBM | 2-D neurogram coefficients | TIMIT | 59.37 |
| | | | | TIDIGT | 35.46 |
| | | | | YOHO | 36.5 |
| | | | | UM (text-dependent) | 20.91 |
| Faragallah [19] | noisy | SVM | MKMFCC | Private Arabic dataset | 3.5 |
| Korba et al. [20] | noisy | GMM | MVA method applied to the MFCC features as post-processing stage | TIMIT dataset | 40.66 |
| Ayhan and Kwan [21] | noisy | GMM | GFCC with Bounded marginalization | Private dataset 1 | 38.06 |
| | | | | Private dataset 2 | 34.32 |
| | | | | RM1 | 4.28 |

### Acknowledgements

"We would like to thank the University of Sharjah for funding this work through the two competitive research projects entitled Capturing, Studying, and Analyzing Arabic Emirati-Accented Speech Database in Stressful and Emotional Talking Environments for Different Applications, No.